\documentclass{article}

\usepackage{xcolor}
\usepackage{tcolorbox}
\tcbuselibrary{breakable,skins,listings}
\usepackage[preprint]{neurips_2026}
\usepackage{amsmath} 
\usepackage{graphicx}
\usepackage{xcolor}
\usepackage{booktabs}
\usepackage{multirow} 

\usepackage[utf8]{inputenc} 
\usepackage[T1]{fontenc}    
\usepackage{hyperref}       
\usepackage{url}            
\usepackage{booktabs}       
\usepackage{amsfonts}       
\usepackage{nicefrac}       
\usepackage{microtype}      
\usepackage{xcolor}         
\title{DCVD: Dual-Channel Cross-Modal Fusion for Joint Vulnerability Detection and Localization}

\author{%
\begin{tabular}{c}
Wenxin Tang\textsuperscript{1,*} \quad
Junliang Liu\textsuperscript{2,*} \quad
Wenbin Li\textsuperscript{3} \quad
Jingyu Xiao\textsuperscript{4} \quad
Xi Xiao\textsuperscript{1} \\
Mingzhe Liu\textsuperscript{5} \quad
Jinlong Yang\textsuperscript{6} \quad
Xuan Liu\textsuperscript{7} \quad
Yuehe Ma\textsuperscript{8} \\
Wang Luo\textsuperscript{9} \quad
Qing Li\textsuperscript{10} \quad
Lei Wang\textsuperscript{11} \quad
Peng Xiangli\textsuperscript{12} \\[0.3em]
\normalfont\small
\textsuperscript{1}Tsinghua University \quad
\textsuperscript{2}Dalian Maritime University \quad
\textsuperscript{3}Hunan University \\
\normalfont\small
\textsuperscript{4}The Chinese University of Hong Kong \quad
\textsuperscript{5}Shenzhen University \\
\normalfont\small
\textsuperscript{6}Northwestern Polytechnical University \quad
\textsuperscript{7}Shandong University \\
\normalfont\small
\textsuperscript{8}BNU-HKBU United International College \quad
\textsuperscript{9}Sun Yat-sen University \\
\normalfont\small
\textsuperscript{10}Peng Cheng Laboratory \quad
\textsuperscript{11}Guangzhou Intelligence Communications Technology Co., Ltd. \\
\normalfont\small
\textsuperscript{12}The Fifth Electronic Research Institute of MIIT \\[0.3em]
\normalfont\small
\texttt{twx24@mails.tsinghua.edu.cn} \quad
\texttt{1120241292ljl@dlmu.edu.cn} \quad
\texttt{xiaox@sz.tsinghua.edu.cn}
\end{tabular}
}

\begin{document}

\maketitle

\begingroup
\renewcommand{\thefootnote}{\fnsymbol{footnote}}
\footnotetext[1]{These authors contributed equally to this work.}
\endgroup

\begin{abstract}
Software vulnerability detection plays a critical role in ensuring system security, where real-world auditing requires not only determining whether a function is vulnerable but also pinpointing the specific lines responsible. However, existing approaches either rely on a single information source---sequential, structural, or semantic---failing to jointly exploit the complementary strengths across modalities, or treat statement-level localization merely as a byproduct of function-level detection without explicit line-level supervision. To address these limitations, we propose DCVD (Dual-Channel Cross-Modal Vulnerability Detection), a unified framework that performs joint function-level detection and statement-level localization. DCVD extracts control-dependency and semantic features through two parallel branches and integrates them via contrastive alignment coupled with bidirectional cross-attention, effectively bridging the cross-modal representation gap. It further introduces explicit supervision signals at both the function and statement levels, enabling collaborative optimization across the two granularities. Extensive experiments on a large-scale real-world vulnerability benchmark demonstrate that DCVD consistently outperforms state-of-the-art methods on both function-level detection and statement-level localization. Our
code is available at ~\url{https://github.com/vinsontang1/DCVD}.
\end{abstract}

\section{Introduction}

The increasing scale and complexity of modern software systems have made security vulnerabilities a pervasive and critical concern. From buffer overflows~\cite{cowan2000buffer,lhee2003buffer} and use-after-free~\cite{caballero2012undangle,lee2015preventing} errors to injection attacks~\cite{zou2025queryattack,ray2012defining} and authentication bypasses~\cite{shoshitaishvili2015firmalice}, software vulnerabilities~\cite{sanchez2019software} pose severe threats to system integrity, user privacy, and economic security. As software development accelerates driven by agile methodologies and open-source ecosystems, the volume of code requiring security review has far outpaced the capacity of manual auditing. Consequently, automated vulnerability detection has emerged as an indispensable capability in the software security lifecycle. Recent advances in deep learning and large language models (LLMs) have demonstrated promising results in software understanding tasks~\cite{liu2025benchmarking,xiao2025efficientuicoder}, opening new avenues for learning-based vulnerability detection. However, real-world security auditing demands not only the ability to determine \textit{whether} a function contains a vulnerability (function-level detection), but also the ability to pinpoint \textit{which specific lines} are responsible (statement-level localization). Simultaneously satisfying both requirements remains an open and challenging problem.

Existing approaches to vulnerability detection can be broadly categorized into three paradigms based on their primary information source. \textbf{(a) Sequence-based methods} model source code as token sequences and leverage Transformer architectures to capture contextual semantics~\cite{liu2024pre,wen2023less,du2024generalization}. For instance, LineVul~\cite{fu2022linevul} applies a pre-trained RoBERTa model to learn token-level representations for vulnerability prediction. While such methods effectively capture sequential semantic patterns, they inherently discard the structural information embedded in program control flow and syntax hierarchies. \textbf{(b) Graph-based methods} take program representations such as Abstract Syntax Trees (ASTs), Control Flow Graphs (CFGs), and Program Dependence Graphs (PDGs) as input to capture structural dependencies within code~\cite{wang2024combining,zhang2023vulnerability,weng2024matsvd,peng2023ptlvd}. Representative works including MatsVD~\cite{weng2024matsvd} and PTLVD~\cite{peng2023ptlvd} have shown that structural information provides valuable signals for vulnerability detection. However, these methods primarily focus on syntactic and control-flow patterns while lacking the ability to comprehend high-level semantic intent and functional logic of the code. \textbf{(c) LLM-based methods} directly employ Code LLMs for vulnerability detection, leveraging their strong semantic understanding capabilities acquired through large-scale pre-training on code corpora~\cite{wen2024scale,yang2024security,du2024generalization}. Despite their powerful language comprehension, these approaches lack explicit modeling of program structural properties, limiting their sensitivity to vulnerabilities rooted in control flow anomalies. These three paradigms each rely on a single information source and fail to effectively integrate structural and semantic information, resulting in limited performance when confronting complex vulnerability patterns that simultaneously involve control flow irregularities and semantic anomalies. Moreover, existing frameworks that attempt both function-level detection and statement-level localization, such as PTLVD~\cite{peng2023ptlvd} and LineVul~\cite{fu2022linevul}, treat statement-level localization merely as a byproduct of function-level detection---typically derived through attention weights or gradient-based attribution---without incorporating explicit statement-level supervision signals, leading to insufficient localization precision.

To advance automated vulnerability detection, three key technical challenges must be addressed: \textbf{(1) How to effectively extract vulnerability-relevant control dependency information.} Vulnerabilities in real-world software often do not originate from isolated statements but arise from complex interactions among control dependencies---a missing boundary check may render a subsequent array access unsafe, or an incorrectly configured conditional branch may expose an unauthorized execution path. These vulnerabilities are deeply embedded within control flow structures, making them difficult to identify through surface-level analysis. Existing static analysis techniques provide coarse-grained representations of control dependencies that lack the flexibility to distinguish, within complex nested structures and conditional jumps, which control paths are critical for vulnerability triggering and which constitute irrelevant noise. \textbf{(2) How to bridge the cross-modal representation gap between structural and semantic features.} Control dependency information is naturally represented as graph structures emphasizing syntactic and flow relationships, while semantic information manifests as continuous embedding vectors capturing functional intent and behavioral logic. These two modalities exhibit fundamentally different data formats, semantic granularities, and feature distributions. Native fusion strategies such as concatenation or weighted averaging fail to achieve meaningful alignment and may introduce representational conflicts that degrade detection performance. \textbf{(3) How to reconcile the supervision asymmetry between function-level detection and statement-level localization.} Function-level detection constitutes a coarse-grained binary classification task, whereas statement-level localization requires fine-grained sequence-level prediction. The two tasks differ substantially in label density and gradient signal strength. Existing methods lack explicit statement-level supervision, relegating localization to an auxiliary output derived indirectly from function-level predictions, which fundamentally constrains localization accuracy.

To address these challenges, we propose \textbf{DCVD} (\textbf{D}ual-\textbf{C}hannel Cross-Modal \textbf{V}ulnerability \textbf{D}etection), a unified framework that synergistically integrates multi-source information extraction, cross-modal feature fusion, and multi-granularity supervised learning for joint vulnerability detection and localization. To tackle \textbf{Challenge (1)}, we propose the \textit{Dual-Channel Encoder}, which comprises a structure branch and a semantic branch operating in parallel: the former employs Graph Attention Networks (GATs) to extract control dependency features from AST and CFG representations, automatically focusing on vulnerability-relevant control nodes and paths; the latter leverages an LLM to generate natural language explanations of the code and extracts deep semantic features through a shared embedding layer. The two branches provide complementary coverage of both structural and semantic dimensions. To address \textbf{Challenge (2)}, we introduce the \textit{Cross-Modal Fusion} module, which first aligns the two modalities into a unified high-dimensional space via contrastive learning, and then employs bidirectional cross-attention to enable deep interaction and mutual enhancement, producing a fused representation that integrates both structural and semantic information. To resolve \textbf{Challenge (3)}, we design the \textit{Multi-Granularity Supervisor}, which comprises function-level and statement-level supervision branches operating in parallel: the former optimizes overall vulnerability detection through classification loss, while the latter incorporates explicit statement-level supervision signals for fine-grained localization. The two branches are jointly optimized to achieve collaborative detection and localization. Additionally, we employ a Transformer-based pre-trained language model to perform deep contextual modeling on the fused features before the supervision module, further enhancing the model's capacity to capture long-range dependencies and cross-line contextual information.

Our main contributions are summarized as follows:
\begin{itemize}
    \item We propose the \textit{Dual-Channel Encoder} that combines GAT-based control dependency feature extraction with LLM-enhanced semantic feature extraction in a parallel architecture, effectively capturing multi-dimensional information closely related to vulnerabilities from both structural and semantic perspectives.
    \item We introduce the \textit{Cross-Modal Fusion} module based on contrastive learning and bidirectional cross-attention, which bridges the representation gap between graph-structural and textual-semantic features and achieves deep complementary fusion of the two information sources.
    \item We design the \textit{Multi-Granularity Supervisor} that incorporates explicit statement-level supervision alongside function-level supervision, enabling collaborative optimization of vulnerability detection and localization without relying on indirect attention-based attribution.
    \item Experimental results demonstrate that our framework consistently outperforms state-of-the-art methods for both function-level detection and statement-level localization.
\end{itemize}

\begin{figure*}
    \centering
    \includegraphics[width=\linewidth]{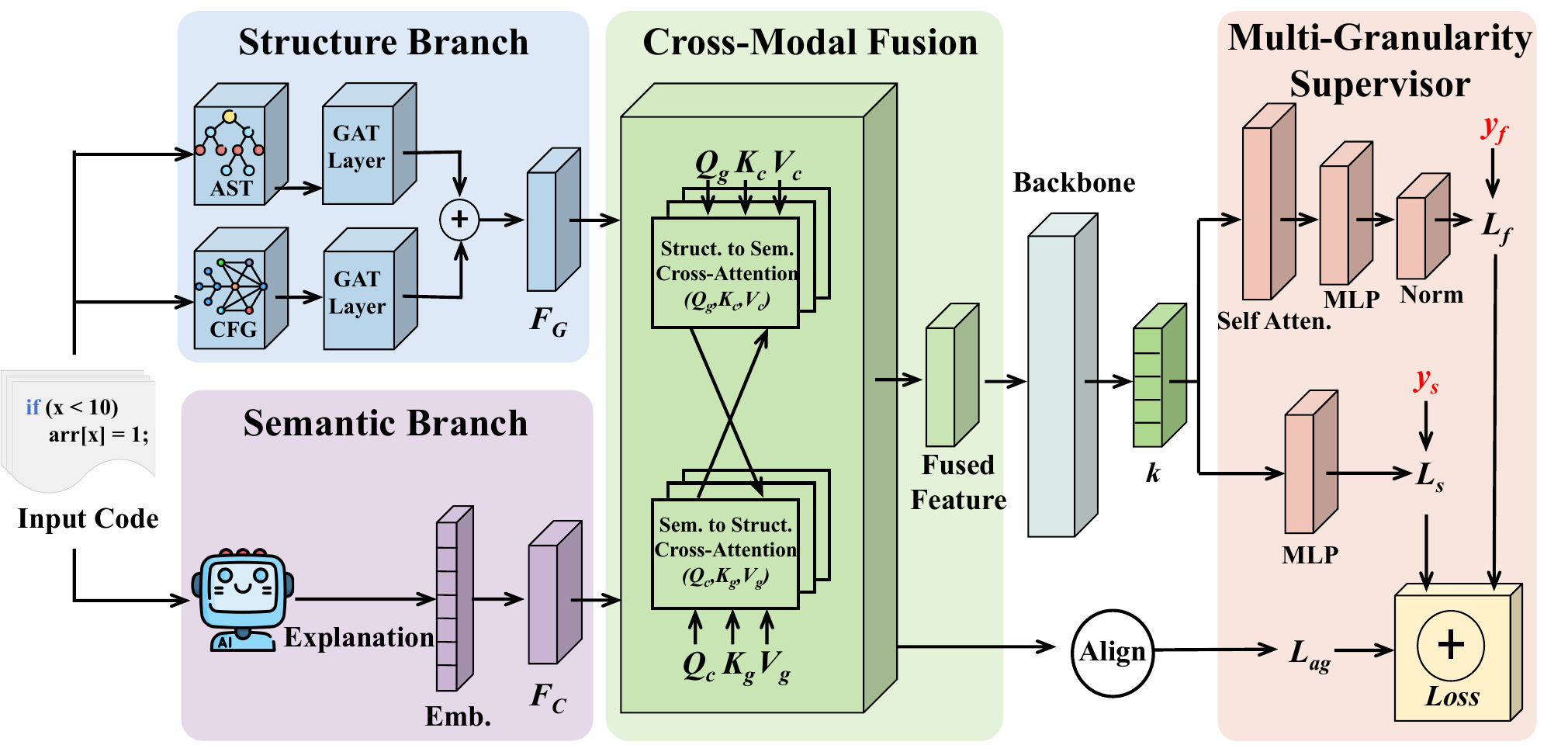}
    \caption{The framework of DCVD.}
    \label{fig:framework}
\end{figure*}

\section{Related Work}

\subsection{Semantic-Based Vulnerability Detection}

With the strong code understanding capabilities demonstrated by large language models~\cite{tang2025slidecoder,tang2026efficientpostergen,luo2026rethinking}, semantic modeling based on pre-trained models and LLMs has been widely applied to code-related tasks~\cite{xiao2025interaction2code,xiao2025efficientuicoder}, including vulnerability detection. Early work treats source code as token sequences and leverages pre-trained Transformers for modeling. LineVul~\cite{fu2022linevul} adopts CodeBERT~\cite{feng2020codebert} with BPE tokenization~\cite{berglund2023formalizing} for function-level vulnerability prediction, and infers line-level results through self-attention weights; VulBERTa~\cite{hanif2022vulberta} performs lightweight RoBERTa~\cite{liu2019roberta} pre-training under a customized tokenization scheme that combines BPE with predefined tokens, achieving strong performance under limited data budgets. Along the direction of semantic enrichment, SCALE~\cite{wen2024scale} attaches natural language comments generated by LLMs onto ASTs to compensate for the limited understanding of complex statements in pre-trained models; Counterfactual randomizes the replacement of user-defined identifiers to generate counterfactual data that mitigates superficial spurious correlations; More recent work further exploits the reasoning capabilities of LLMs: VulLLM~\cite{du2024generalization} employs multi-task instruction tuning on detection, localization, and explanation to force LLMs to learn the root causes of vulnerabilities; MSIVD~\cite{yang2024security} jointly trains an LLM fine-tuned with multi-turn self-instructed dialogues together with a data-flow GNN adapter. Nevertheless, these methods primarily rely on semantics at the token or natural language level, and remain limited in modeling the control and data dependencies inherent to code.

\subsection{Structure-Aware Vulnerability Detection}

Another line of research explicitly injects code structure or execution dependencies into models to capture vulnerability patterns that are difficult to characterize through semantics alone. Along the direction of graph and dependency modeling, DFEPT~\cite{jiang2024dfept} extracts data-flow graphs from ASTs and augments pre-trained models with GNNs as a plug-and-play component; PDBERT~\cite{liu2024pre} introduces control-dependency and data-dependency prediction as new pre-training objectives; MatsVD~\cite{weng2024matsvd} converts the dependency relations in PDGs into attention masks that constrain Transformers to attend only among dependent statements. For fine-grained output, PTLVD~\cite{peng2023ptlvd} combines program slicing with Transformers and aggregates token-level contributions from Integrated Gradients and Saliency into line-level scores. Despite the progress in either structural modeling or fine-grained output, most of these methods are confined to a single information source, or treat statement-level localization as a post-hoc explanation of function-level predictions, lacking deep integration of structural and semantic information as well as explicit line-level supervision.

\section{Methodology}
\label{sub:method}
\subsection{Overview}

To jointly achieve function-level vulnerability detection and statement-level localization, we propose the DCVD framework, as illustrated in Figure~\ref{fig:framework}. Given an input function, the Dual-Channel Encoder first extracts code features from both structural and semantic perspectives in parallel, producing $F^{(s)}$ and $F^{(t)}$ respectively. The Cross-Modal Fusion module then aligns the two representations via contrastive learning and fuses them through bidirectional cross-attention to obtain the fused feature $H$. A multi-layer Transformer subsequently performs deep contextual modeling on $H$, capturing long-range dependencies and cross-line contextual information to yield the enriched representation $K$. Finally, the Multi-Granularity Supervisor conducts joint supervised training on $K$ for both function-level detection and statement-level localization.

\subsection{Dual-Channel Encoder}

\textbf{Structure Branch.} The structure branch captures control dependency information critical for identifying vulnerabilities rooted in control flow anomalies. We employ static analysis tools to extract two complementary graph representations from the source code: the AST encoding hierarchical syntactic structure, and the CFG capturing execution flow and branching logic. Both share the same node set but differ in edge connectivity.

Given node features $X \in \mathbb{R}^{N \times d}$ obtained through a learnable embedding layer, we process the AST and CFG through two independent GAT~\cite{velivckovic2017graph} networks. For each GAT, the attention coefficient between node $i$ and its neighbor $j \in \mathcal{N}(i)$ is:
\begin{equation}
    \alpha_{ij} = \frac{\exp\!\Big(\rho\big(a^{\top}[Wx_i \| Wx_j]\big)\Big)}{\sum_{k \in \mathcal{N}(i)} \exp\!\Big(\rho\big(a^{\top}[Wx_i \| Wx_k]\big)\Big)}
\end{equation}
where $W$ is a learnable weight matrix, $a$ is the attention parameter vector, $\|$ denotes concatenation, and $\rho$ is a nonlinear activation function. The updated representation of node $i$ is obtained by aggregating neighbor features weighted by the attention coefficients:
\begin{equation}
    x_i' = \sigma\!\left(\sum_{j \in \mathcal{N}(i)} \alpha_{ij}\, Wx_j\right)
\end{equation}
where $\sigma$ denotes an activation function. Through this mechanism, the GAT adaptively assigns higher weights to control nodes most relevant to vulnerability patterns, such as boundary checks and conditional branches, while suppressing noise from irrelevant structural elements.

Let $f_A$ and $f_C$ denote the two GAT networks operating on the AST and CFG respectively. The structural representation is obtained by element-wise addition:
\begin{equation}
    F^{(s)} = f_A(X, E_{A}) + f_C(X, E_{C})
\end{equation}
where $E_{A}$ and $E_{C}$ are the edge index matrices. The resulting $F^{(s)} \in \mathbb{R}^{N \times d'}$ encodes both syntactic hierarchies and control flow dependencies.

\textbf{Semantic Branch.} The semantic branch extracts features capturing functional intent, behavioral logic, and API call semantics that are difficult to obtain from structural analysis alone. We first feed the preprocessed source code into an LLM to generate a natural language explanation describing the code's functionality and potential security-relevant behaviors. This explanation surfaces implicit logic and external API semantics not directly expressed in the code syntax.

Both the source code and its explanation are tokenized and passed through a shared embedding layer $\mathcal{E}$ extracted from a pre-trained language model. Let $C = \mathcal{E}(c) \in \mathbb{R}^{M \times d_h}$ and $T = \mathcal{E}(t) \in \mathbb{R}^{L \times d_h}$ denote the code and explanation embeddings respectively, where $M$ and $L$ are the sequence lengths. We compute a global explanation vector via masked mean pooling:
\begin{equation}
    \bar{t} = \frac{\sum_{j=1}^{L} T_j \cdot m_j}{\sum_{j=1}^{L} m_j}
\end{equation}
where $m_j$ is the attention mask indicating valid tokens. The semantic representation is obtained by injecting the global explanation into each code token and projecting to the target dimension: $F^{(t)} = \phi\!\left(C + \bar{t}\right)$, where $\phi$ is a nonlinear projection mapping to dimension $d'$. The resulting $F^{(t)} \in \mathbb{R}^{M \times d'}$ incorporates both token-level code semantics and the global functional understanding provided by the LLM explanation.

\subsection{Cross-Modal Fusion}

The Dual-Channel Encoder produces two complementary yet heterogeneous representations: $F^{(s)}$ from graph neural networks and $F^{(t)}$ from language model embeddings. These modalities reside in different representation subspaces, and direct concatenation would fail to exploit their complementary nature. We address this through a two-stage strategy: contrastive alignment followed by bidirectional cross-attention.

\textbf{Contrastive Alignment.} We first align the two modalities into a shared space through contrastive learning. For each sample $i$ in a mini-batch of size $B$, we derive global representations $g_i$ from $F^{(s)}$ via mean pooling over graph nodes, and $c_i$ from the semantic branch. The pair $(g_i, c_i)$ from the same function constitutes a positive pair, while cross-function pairs serve as negatives. The alignment loss is

\begin{equation}
    \mathcal{L}_{align} = -\frac{1}{B}\sum_{i=1}^{B} \log \frac{\exp(\mathrm{sim}(g_i, c_i) / \tau)}{\sum_{j=1}^{B} \exp(\mathrm{sim}(g_i, c_j) / \tau)}
\end{equation}

where $\mathrm{sim}(\cdot,\cdot)$ denotes cosine similarity and $\tau$ is a temperature hyperparameter. This objective pulls same-function representations together while pushing different-function representations apart, reducing the cross-modal distribution gap before fusion.

\textbf{Bidirectional Cross-Attention.} After alignment, we employ bidirectional cross-attention to enable deep interaction between the two modalities. Unlike self-attention, cross-attention derives queries from one modality and keys/values from the other, allowing each modality to selectively attend to the most informative elements of its counterpart.

In the structure-to-semantic direction, the structural representation queries the semantic feature: $H_s = \mathrm{softmax}\!\left(Q_s {K_t}^{\!\top} / \sqrt{d'}\right) V_t$, where $Q_s = F^{(s)}W_Q^{s}$, $K_t = F^{(t)}W_K^{t}$, and $V_t = F^{(t)}W_V^{t}$ are linear projections. In the semantic-to-structure direction: $H_t = \mathrm{softmax}\!\left(Q_t {K_s}^{\!\top} / \sqrt{d'}\right) V_s$, where $Q_t = F^{(t)}W_Q^{t}$, $K_s = F^{(s)}W_K^{s}$, and $V_s = F^{(s)}W_V^{s}$. Since $H_s \in \mathbb{R}^{N \times d'}$ operates at the graph-node level while $H_t \in \mathbb{R}^{M \times d'}$ operates at the token level, we apply mean pooling to $H_s$ and broadcast the resulting vector to match the token sequence length. The fused feature is then:
\begin{equation}
    H = \sigma\!\left(W_m \left[H_t \| \bar{H}_s\right]\right)
\end{equation}
where $W_m \in \mathbb{R}^{d' \times 2d'}$ is a learnable projection, $\bar{H}_s$ denotes the broadcast vector, and $\sigma$ is an activation function. The resulting $H \in \mathbb{R}^{M \times d'}$ enables each token position to be informed by both control dependency context and semantic intent through mutual querying.

After that, we employ a multi-layer Transformer to perform deep contextual modeling on $H$, capturing long-range dependencies and cross-line contextual patterns. Specifically, the fused embeddings are projected to match the hidden dimension and fed directly into the Transformer encoder, yielding $K = f_{\mathrm{T}}(W_{a}\, H)$, where $W_{a}$ is a dimension alignment projection and $f_{\mathrm{T}}$ denotes the multi-layer Transformer. The output $K \in \mathbb{R}^{M \times d_k}$ serves as the shared input to both branches of the Multi-Granularity Supervisor.

\subsection{Multi-Granularity Supervisor}

To jointly optimize vulnerability detection and localization under a unified framework, we design the Multi-Granularity Supervisor, which comprises two parallel branches: a function-level detection branch and a statement-level localization branch, each equipped with its own supervision signal.

\textbf{Function-Level Detection.} The function-level branch leverages function-level labels to determine whether the input function contains a vulnerability. The sequence representation $K$ is pooled into a global vector and passed through an MLP $g_f$ to produce a binary prediction:
\begin{equation}
    \hat{y}_f = \sigma\!\left(g_f\!\left(\bar{K}\right)\right)
\end{equation}
where $\bar{K}$ denotes the pooled representation and $\sigma$ is the sigmoid function. The function-level loss $\mathcal{L}_f$ is computed using binary cross-entropy between $\hat{y}_f$ and the ground-truth label $y_f$.

\textbf{Statement-Level Localization.} The statement-level branch leverages line-level labels to identify the specific lines responsible for the vulnerability. We first refine the token-level representations through a self-attention layer $f_{sa}$ followed by an MLP $g_s$ and layer normalization:
\begin{equation}
    \tilde{K} = \mathrm{LN}\!\left(K + g_s\!\left(f_{sa}(K)\right)\right)
\end{equation}
Each token representation is then projected to a scalar vulnerability score. Since labels are annotated at the line level, we aggregate token scores within each line by averaging:
\begin{equation}
    s_l = \frac{1}{|\mathcal{T}_l|}\sum_{i \in \mathcal{T}_l} w^{\top} \tilde{K}_i
\end{equation}
where $\mathcal{T}_l$ is the set of token indices belonging to line $l$ and $w$ is a learnable projection vector. The line-level vulnerability probability is $\hat{y}_s^{(l)} = \sigma(s_l)$. The statement-level loss $\mathcal{L}_s$ is computed via KL divergence between the predicted and ground-truth line-level distributions.

After obtaining both branch losses, we combine them with the contrastive alignment loss to form the total training objective:
\begin{equation}
    \mathcal{L} = \alpha \left(\mathcal{L}_f + \beta \cdot \mathcal{L}_{align}\right) + (1 - \alpha) \cdot \mathcal{L}_s
\end{equation}
where $\alpha$ and $\beta$ are hyperparameters balancing the contributions. The function-level branch provides global guidance on vulnerable regions, the statement-level branch delivers fine-grained localization through explicit supervision, and the alignment loss ensures consistent cross-modal representations throughout training.

\section{Experiments}

\subsection{Experimental Setup}\label{subsec:expset}

\textbf{Dataset.} We conduct experiments on BigVul~\cite{fan2020ac}, a large-scale real-world C/C++ vulnerability dataset that provides both function-level vulnerability labels and statement-level annotations indicating the specific lines responsible for each vulnerability.

\textbf{Baselines.} We compare DCVD against seven representative vulnerability detection methods spanning three paradigms. (1) \textbf{TextCNN}~\cite{guo2019improving} applies multi-kernel convolutions over token sequences for sequence-level classification. (2) \textbf{LineVul}~\cite{fu2022linevul} fine-tunes a Transformer-based pre-trained model on source code tokens and derives statement-level predictions from attention weights. (3) \textbf{IVDetect}~\cite{li2021vulnerability} employs a feature-attention GCN over program dependence graphs enriched with multi-type contextual features. (4) \textbf{LineVD}~\cite{hin2022linevd} formulates statement-level localization as a node classification task on statement-level graphs. (5) \textbf{VELVET}~\cite{ding2022velvet} ensembles a Transformer and a gated graph neural network to jointly exploit sequential and structural information. (6) \textbf{ICVH}~\cite{nguyen2021information} combines code representation learning with a hierarchical classification architecture for vulnerability identification. (7) \textbf{MatsVD}~\cite{weng2024matsvd} adopts a multi-task architecture that jointly performs function-level detection and statement-level localization.

\textbf{Evaluation Metrics.} We evaluate DCVD from two complementary perspectives. For \textit{classification quality}, we report Matthews Correlation Coefficient (Mcc), Precision (Pre), Recall (Re), and F1-score at both the function level and the statement level. Statement-level classification is further evaluated under two protocols: the \textit{Two-Phase} setting, where statement-level prediction is conducted only on functions predicted as vulnerable by the function-level branch, and the \textit{One-Phase} setting, where statement-level prediction is conducted independently of the function-level branch. For \textit{localization quality}, we adopt Top-1, Top-3, and Top-5 Accuracy, Mean First Rank (MFR), and Mean Average Rank (MAR), which jointly characterize how precisely a method ranks vulnerable lines among all lines in a function.

The semantic branch of DCVD uses GPT-4o-mini-2024-07-18 to produce natural language explanations of the input code, and the multi-layer Transformer for deep contextual modeling is instantiated as GraphCodeBERT~\cite{guo2020graphcodebert}. The entire framework is trained end-to-end on NVIDIA H20 GPUs.

\subsection{Main Results}

Table~\ref{tab:main_cls} presents the classification performance of different methods at both function and statement levels, and Table~\ref{tab:main_rank} presents the statement-level localization ranking performance, evaluated using the metrics introduced in Section~\ref{subsec:expset}.

\begin{table*}[t]
\centering
\caption{Function-level and statement-level classification performance. Statement-level results are reported under both Two-Phase and One-Phase protocols.}
\label{tab:main_cls}
\small
\setlength{\tabcolsep}{4pt}
\resizebox{\textwidth}{!}{
\begin{tabular}{l cccc cccc cccc}
\toprule
\multirow{2}{*}{Method} & \multicolumn{8}{c}{Two-Phase Detection} & \multicolumn{4}{c}{One-Phase Detection} \\
\cmidrule(lr){2-9} \cmidrule(lr){10-13}
 & \multicolumn{4}{c}{Function Level} & \multicolumn{4}{c}{Statement Level} & \multicolumn{4}{c}{Statement Level} \\
\cmidrule(lr){2-5} \cmidrule(lr){6-9} \cmidrule(lr){10-13}
 & Mcc & Pre & Re & F1 & Mcc & Pre & Re & F1 & Mcc & Pre & Re & F1 \\
\midrule
TextCNN   & 0.3849 & 61.75 & 26.20 & 36.79 & 0.1404 & 21.10 & 30.15 & 24.83 & 0.2206 & 26.24 & 40.27 & 31.78 \\
ICVH      & 0.4556 & 65.73 & 33.94 & 44.76 & 0.1804 & 21.39 & 43.85 & 28.75 & 0.2656 & 27.34 & 50.76 & 35.54 \\
IVDetect  & 0.6171 & 72.28 & 55.23 & 62.61 & 0.3014 & 28.36 & 58.46 & 38.19 & 0.3564 & 31.52 & 65.06 & 42.47 \\
LineVD    & 0.7537 & 82.47 & 70.76 & 76.17 & 0.3899 & 35.52 & 63.86 & 45.65 & 0.4506 & 40.94 & 67.38 & 50.93 \\
LineVul   & 0.8224 & 87.65 & 78.58 & 82.87 & 0.4951 & 62.10 & 47.45 & 53.79 & 0.5299 & 62.38 & 53.40 & 57.54 \\
VELVET    & 0.8973 & 95.41 & 85.19 & 90.01 & 0.6966 & 68.74 & 71.23 & 69.96 & 0.6781 & 59.87 & \textbf{86.25} & 70.68 \\
MatsVD    & 0.9361 & \textbf{97.53} & 90.37 & 93.81 & 0.8606 & 91.57 & 81.13 & 86.03 & 0.8585 & 89.90 & 84.86 & 87.31 \\
\midrule
\textbf{DCVD (ours)} & \textbf{0.9467} & 97.50 & \textbf{92.32} & \textbf{94.84} & \textbf{0.8824} & \textbf{92.59} & \textbf{84.18} & \textbf{88.18} & \textbf{0.8791} & \textbf{92.70} & 84.78 & \textbf{88.56} \\
\bottomrule
\end{tabular}
}
\end{table*}

\begin{table}[t]
\centering
\caption{The statement-level localization ranking performance.}
\label{tab:main_rank}
\small
\begin{tabular}{l ccccc}
\toprule
Method & Top-1 & Top-3 & Top-5 & MFR & MAR \\
\midrule
TextCNN  & 27.41 & 34.79 & 46.01 & 13.16 & 18.69 \\
ICVH     & 35.37 & 43.08 & 55.83 & 11.80 & 16.24 \\
IVDetect & 46.69 & 54.27 & 62.97 &  8.73 & 12.63 \\
LineVD   & 60.52 & 73.76 & 80.74 &  4.63 & 10.21 \\
LineVul  & 24.54 & 48.49 & 60.64 &  9.32 & 13.94 \\
VELVET   & 86.09 & 92.41 & 93.86 &  3.73 &  7.81 \\
MatsVD   & 92.77 & 94.46 & 94.82 &  2.46 &  6.34 \\
\midrule
\textbf{DCVD (ours)} & \textbf{93.80} & \textbf{97.01} & \textbf{97.86} & \textbf{1.77} & \textbf{4.17} \\
\bottomrule
\end{tabular}
\end{table}

DCVD outperforms the baselines on nearly all metrics. For function-level detection, the largest gains over the strongest baseline MatsVD are observed on F1 and Recall, which improve from 93.81 and 90.37 to 94.84 and 92.32 respectively (+1.03 and +1.95), confirming that integrating complementary structural and semantic information through the Dual-Channel Encoder and Cross-Modal Fusion effectively overcomes the limitations of methods relying on a single information source.

Benefiting from the explicit statement-level supervision introduced by the Multi-Granularity Supervisor, DCVD also leads on statement-level classification under both the Two-Phase and One-Phase protocols. Under Two-Phase, Mcc and F1 improve from 0.8606 and 86.03 to 0.8824 and 88.18 (+0.0218 and +2.15); under One-Phase, they improve from 0.8585 and 87.31 to 0.8791 and 88.56 (+0.0206 and +1.25). The consistent gains across both protocols indicate that the statement-level branch of DCVD is not merely a byproduct of the function-level prediction but possesses independent discriminative capability.

On the ranking metrics, DCVD achieves the best performance across Top-1/3/5, MFR, and MAR. In particular, Top-1 increases from 92.77 to 93.80 (+1.03), MFR decreases from 2.46 to 1.77 ($-0.69$ positions), and MAR decreases from 6.34 to 4.17 ($-2.17$ positions). This demonstrates that DCVD ranks truly vulnerable lines substantially closer to the top, rather than relying on indirect attention- or gradient-based attribution as in prior work.

Overall, the results across classification and ranking metrics show that DCVD consistently improves both function-level detection and statement-level localization over existing methods on BigVul.

\subsection{Ablation Study}
To validate each component of DCVD, we design four ablation variants. \textbf{w/o Structure} removes the entire structure branch, relying solely on the semantic branch. \textbf{w/o Semantic} removes the semantic branch and retains only the structural branch. \textbf{w/o Fusion} replaces the contrastive alignment and bidirectional cross-attention with simple concatenation followed by an MLP. \textbf{w/o Multi-Task} removes the statement-level supervision, reducing the Multi-Granularity Supervisor to a single function-level branch (function-level only). We randomly sample 30\% of the test set for evaluation, with classification and ranking results reported in Table~\ref{tab:ablation_cls} and Table~\ref{tab:ablation_rank}.

\begin{table*}[t]
\centering
\caption{The function-level and statement-level classification performance of ablation study.}
\label{tab:ablation_cls}
\small
\setlength{\tabcolsep}{4pt}
\resizebox{\textwidth}{!}{
\begin{tabular}{l cccc cccc cccc}
\toprule
\multirow{2}{*}{Setting} & \multicolumn{8}{c}{Two-Phase Detection} & \multicolumn{4}{c}{One-Phase Detection} \\
\cmidrule(lr){2-9} \cmidrule(lr){10-13}
 & \multicolumn{4}{c}{Function Level} & \multicolumn{4}{c}{Statement Level} & \multicolumn{4}{c}{Statement Level} \\
\cmidrule(lr){2-5} \cmidrule(lr){6-9} \cmidrule(lr){10-13}
 & Mcc & Pre & Re & F1 & Mcc & Pre & Re & F1 & Mcc & Pre & Re & F1 \\
\midrule
\textbf{Ours}  & \textbf{0.9360} & 96.55 & \textbf{91.30} & \textbf{93.85} & \textbf{0.8736} & \textbf{91.68} & \textbf{83.36} & \textbf{87.32} & \textbf{0.8690} & \textbf{91.79} & \textbf{83.94} & \textbf{87.69} \\
w/o Structure  & 0.8774 & 95.65 & 81.57 & 88.05 & 0.6256 & 74.40 & 53.05 & 61.94 & 0.6129 & 77.15 & 53.28 & 63.03 \\
w/o Semantic   & 0.8682 & 94.52 & 80.92 & 87.19 & 0.5861 & 74.56 & 46.50 & 57.28 & 0.5794 & 78.96 & 46.67 & 58.66 \\
w/o Fusion     & 0.8405 & \textbf{96.60} & 74.37 & 84.04 & 0.5493 & 83.45 & 36.46 & 50.74 & 0.5333 & 84.71 & 36.77 & 51.28 \\
w/o Multi-Task & 0.8399 & 89.35 & 80.48 & 84.68 & -- & -- & -- & -- & -- & -- & -- & -- \\
\bottomrule
\end{tabular}
}
\end{table*}

\begin{table}[t]
\centering
\caption{The statement-level localization ranking performance of ablation study.}
\label{tab:ablation_rank}
\small
\begin{tabular}{l ccccc}
\toprule
Setting & Top-1 & Top-3 & Top-5 & MFR & MAR \\
\midrule
\textbf{Ours}  & \textbf{92.87} & \textbf{96.65} & \textbf{97.48} & \textbf{1.80} & \textbf{4.21} \\
w/o Structure  & 80.77 & 87.66 & 89.53 & 7.90 & 12.10 \\
w/o Semantic   & 79.20 & 85.51 & 88.52 & 8.04 & 12.42 \\
w/o Fusion     & 78.91 & 85.08 & 88.09 & 8.06 & 12.49 \\
w/o Multi-Task & -- & -- & -- & -- & -- \\
\bottomrule
\end{tabular}
\end{table}

After removing each component, the performance exhibits varying degrees of decline across function-level and statement-level metrics, demonstrating the contribution of each module to the overall framework. In particular, \textbf{w/o Fusion} causes the largest drop among the feature-related variants on statement-level classification, confirming that naive concatenation is insufficient to bridge the heterogeneous subspaces of graph-structural and textual-semantic representations. More notably, after removing the statement-level supervision signal in \textbf{w/o Multi-Task}, the function-level metrics also drop substantially, indicating that statement-level supervision not only enables fine-grained localization but also provides auxiliary signals that reinforce function-level classification---thereby validating the synergistic design of the Multi-Granularity Supervisor.

\subsection{Parameter Analysis}

\begin{figure*}
    \centering
    \includegraphics[width=\linewidth]{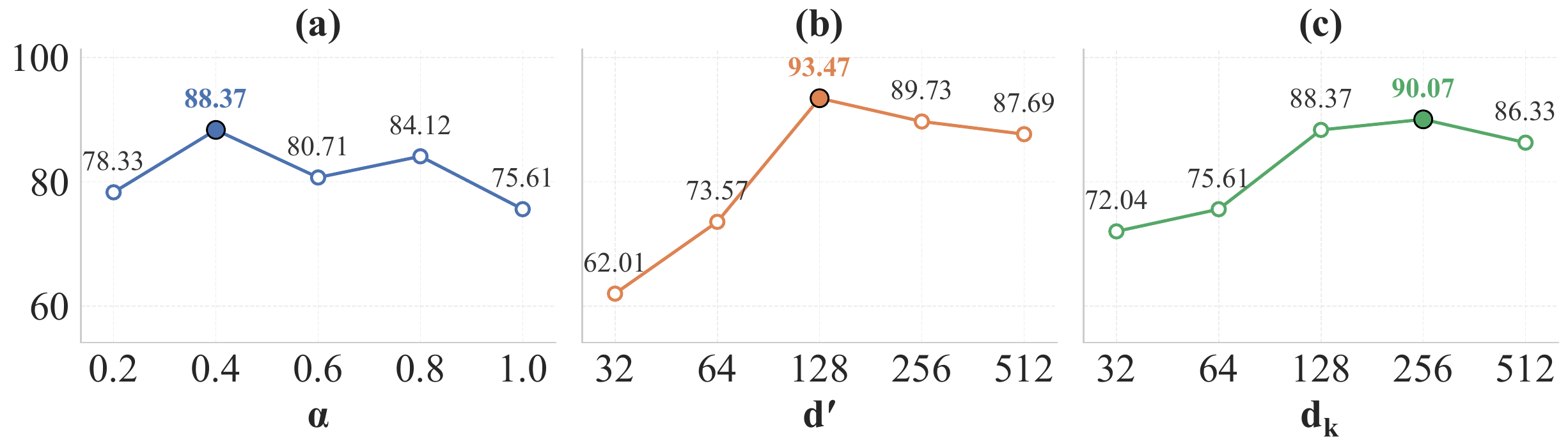}
    \caption{Parameter sensitivity analysis of DCVD. We report the composite Score under different values of (a) the loss-balancing coefficient $\alpha$, (b) the branch/fusion dimension $d'$, and (c) the deep representation dimension $d_k$.}
    \label{fig:param}
\end{figure*}

We further investigate the impact of three key hyperparameters on DCVD, including $\alpha$ (the balancing coefficient between function-level and statement-level losses), $d'$ (the output dimension of both branches and the fused representation), and $d_k$ (the deep representation dimension produced by the multi-layer Transformer). For ease of visualization, we aggregate four representative metrics into a composite $\text{Score} = (\text{Two-Phase Func F1} + \text{Two-Phase Stmt F1} + \text{One-Phase Stmt F1} + \text{Top-1})/4$. Consistent with the ablation study, the parameter analysis is conducted on a 30\% sampled test set, and the results are shown in Figure~\ref{fig:param}.

As illustrated in the figure, the Score peaks when $\alpha = 0.4$; a smaller $\alpha$ provides insufficient function-level supervision. For $d'$, the Score reaches its maximum of 93.47 at $d' = 128$: too small a dimension leaves the two branches and the fused representation without enough capacity to carry cross-modal information, while too large a dimension introduces redundant parameters and exacerbates overfitting. Regarding $d_k$, the best Score of 90.07 is obtained at $d_k = 256$, as a smaller $d_k$ cannot adequately accommodate the long-range dependencies and cross-line context captured by the multi-layer Transformer, whereas a larger $d_k$ comes with an increased risk of overfitting.

\section{Conclusion}

In this paper, we propose DCVD, a unified framework for joint function-level vulnerability detection and statement-level localization. DCVD extracts complementary structural and semantic representations through a Dual-Channel Encoder, bridges the cross-modal gap via contrastive alignment coupled with bidirectional cross-attention in the Cross-Modal Fusion module, and introduces a Multi-Granularity Supervisor that explicitly supervises both granularities to enable collaborative optimization. Two core insights underpin DCVD: deep cross-modal fusion between structural and semantic representations, and explicit supervision at both the function and statement levels, which together address the limitations of single-source modeling and attention-based line-level attribution. Comprehensive evaluations across classification, ranking, ablation, and parameter analyses consistently validate the effectiveness of each design choice. Extensive experiments on BigVul demonstrate that DCVD consistently outperforms state-of-the-art methods.

\bibliography{dd}
\bibliographystyle{plainnat}



\appendix   
\section{Limitations}\label{app:lim}

While DCVD achieves substantial improvements over existing methods, several limitations remain. First, the semantic branch relies on an offline closed-source LLM (GPT-4o-mini) to generate natural language explanations of code, which introduces API costs, reproducibility concerns, and sensitivity to both the underlying LLM's capability and prompt engineering; exploring open-source code LLMs or end-to-end learnable explanation modules could mitigate this dependency. Second, our evaluation is conducted solely on BigVul, a C/C++ vulnerability dataset, and the generalization of DCVD to other programming languages (e.g., Java, Python, Go) and other benchmarks (e.g., Devign, DiverseVul) has not yet been fully validated. Third, the structure branch depends on static analysis tools to extract ASTs and CFGs, and therefore degrades when the input is an incomplete code fragment (e.g., a GitHub diff hunk or a partially visible function in an IDE) for which static analyzers fail to produce reliable graphs; more robust structural representations or integration with pre-trained dependency predictors may help alleviate this issue.

\section{Implementation Details}

\subsection{Configuration Parameters}\label{app:config}

We summarize the configuration of DCVD in Table~\ref{tab:config}.

\begin{table}[h]
\centering
\caption{Configuration of DCVD.}
\label{tab:config}
\small
\begin{tabular}{l l c}
\toprule
Category & Parameter & Value \\
\midrule
\multirow{6}{*}{Model Architecture}
 & $d'$ (branch / fusion dimension) & 128 \\
 & $d_k$ (deep representation dimension) & 256 \\
 & GAT layers & 2 \\
 & Self-attention heads (statement branch) & 8 \\
 & Max sequence length & 512 \\
\midrule
\multirow{7}{*}{Training}
 & Optimizer & AdamW \\
 & Learning rate & $2 \times 10^{-5}$ \\
 & LR scheduler & Cosine with restarts \\
 & Warmup steps & 500 \\
 & Batch size & 32 \\
 & Epochs & 50 \\
 & $\alpha$ (loss balancing) & 0.4 \\
\bottomrule
\end{tabular}
\end{table}

\subsection{Runtime Environment and Training Time}\label{app:runtime}

DCVD is implemented in PyTorch with PyTorch Geometric for graph operations, and is trained using HuggingFace Transformers and the Accelerate library. All experiments are conducted on a single NVIDIA H20 GPU, and one complete training run takes approximately 20 hours.

\section{Significance Analysis of Ablation Study}\label{Significance}

To verify that the improvements reported in the ablation study are not due to random fluctuations, we further conduct a significance analysis. Specifically, for each variant we independently repeat training three times with different random seeds on the 30\% sampled test set, and perform a paired $t$-test against the full DCVD on both Two-Phase Function-Level F1 and One-Phase Statement-Level F1. The resulting $p$-values are reported in Table~\ref{tab:significance}, where $^{***}$, $^{**}$, and $^{*}$ denote $p < 0.001$, $p < 0.01$, and $p < 0.05$ respectively, and ``n.s.'' indicates non-significance ($p \geq 0.05$).

\begin{table}[h]
\centering
\caption{Paired $t$-test $p$-values of each ablation variant compared with the full DCVD.}
\label{tab:significance}
\small
\begin{tabular}{l cc}
\toprule
Variant & Func F1 ($p$-value) & Stmt F1 ($p$-value) \\
\midrule
w/o Structure  & $0.082$ (n.s.) & $<0.01$~$^{**}$ \\
w/o Semantic   & $<0.05$~$^{*}$ & $<0.05$~$^{*}$ \\
w/o Fusion     & $<0.001$~$^{***}$ & $<0.001$~$^{***}$ \\
w/o Multi-Task & $<0.001$~$^{***}$ & -- \\
\bottomrule
\end{tabular}
\end{table}

Overall, six out of the seven comparisons yield statistically significant differences, confirming that the contributions of the Cross-Modal Fusion module and the Multi-Granularity Supervisor, as well as the overall design of the Dual-Channel Encoder, are robust rather than coincidental. The only non-significant case, namely the function-level F1 of w/o Structure, suggests that the semantic branch alone already provides sufficient signals for coarse-grained function-level classification, while its contribution becomes statistically evident only when the task is extended to the finer statement-level granularity.

\section{Broader Impacts}
\label{sec:broader_impacts}
This work aims to improve automated software vulnerability detection and localization, with the primary goal of assisting developers, security analysts, and maintainers in identifying security-critical code regions more accurately and efficiently. By jointly predicting whether a function is vulnerable and localizing the responsible statements, DCVD can potentially reduce the manual effort required for code auditing, accelerate vulnerability triage, and support the development of more secure software systems. These benefits are particularly relevant for large-scale open-source and industrial codebases, where manual security review is often costly and difficult to scale.

At the same time, vulnerability detection and localization techniques may have potential dual-use risks. In particular, fine-grained localization results could be misused by malicious actors to identify exploitable code regions more efficiently. In addition, incorrect predictions may introduce practical risks: false negatives may leave vulnerabilities undetected, while false positives may increase the workload of developers and security teams. Therefore, DCVD should be used as an assistive tool for defensive security analysis rather than as a replacement for expert review, manual validation, or secure development practices.

To mitigate these risks, we recommend that released artifacts and experimental results be used only for defensive security research, vulnerability mitigation, and responsible software maintenance. When deployed in real-world auditing workflows, predictions should be reviewed by qualified security practitioners before any security-critical decision is made. We also encourage future work to incorporate uncertainty estimation, human-in-the-loop validation, and responsible disclosure mechanisms to further reduce the risk of misuse and improve the reliability of automated vulnerability analysis.

\definecolor{PromptBack}{RGB}{248,248,248}
\definecolor{PromptFrame}{RGB}{210,210,210}

\newtcblisting{promptbox}[2][]{
  enhanced,
  breakable,
  colback=PromptBack,
  colframe=PromptFrame,
  boxrule=0.6pt,
  arc=1.5mm,
  left=1.5mm,
  right=1.5mm,
  top=1mm,
  bottom=1mm,
  title={#2},
  fonttitle=\bfseries,
  listing only,
  listing options={
    basicstyle=\ttfamily\small,
    breaklines=true,
    columns=fullflexible,
    keepspaces=true,
    showstringspaces=false
  },
  #1
}

\section{Prompt Template}
\label{app:prompt_template}

This section provides the prompt template used by the LLM-based semantic branch in DCVD. The placeholder enclosed by angle brackets is replaced with the corresponding input code during inference.

\subsection{Semantic Branch: Code Explanation Generation}
\label{app:prompt_semantic_branch}

The semantic branch invokes an LLM to generate a natural-language explanation of the input code. The generated explanation is used to capture functional intent, behavioral logic, API semantics, and security-relevant behaviors that are difficult to obtain from structural analysis alone.

\begin{promptbox}{Prompt Template for Code Explanation Generation}
You are an expert software security and program analysis assistant.

Given the following source code, generate a concise natural-language explanation of its functionality and potential security-relevant behaviors. The explanation will be used as semantic input for a vulnerability detection and localization model, so it should focus on observable program behavior rather than unsupported speculation.

Please analyze the code from the following aspects:

1. Functionality:
Describe the main purpose of the function and what task it performs.

2. Inputs and Outputs:
Identify important parameters, external inputs, return values, and output behaviors.

3. Control Logic:
Describe key branches, loops, conditions, and execution paths that affect the behavior of the function.

4. Data Flow:
Describe how important data values propagate through the function, especially from inputs to memory operations, pointer operations, array accesses, file operations, system calls, or other sensitive operations.

5. Security-Relevant Behaviors:
Identify potential security-relevant operations, such as boundary checks, null checks, memory allocation or release, pointer dereference, buffer access, arithmetic computation, authentication checks, and error handling.

6. Potential Risk Indicators:
Mention suspicious or risky behaviors only if they are directly observable from the code. Do not overclaim vulnerability existence.

Source Code:
<Code>

Output format:
Functionality: <brief description>
Inputs and Outputs: <brief description>
Control Logic: <brief description>
Data Flow: <brief description>
Security-Relevant Behaviors: <brief description>
Potential Risk Indicators: <brief description or N/A>
\end{promptbox}

\end{document}